\documentclass[pre,preprint,showpacs,floatfix,superscriptaddress]{revtex4}
\usepackage{graphicx,amsmath}
\usepackage{undertilde}

\newcommand*{\GL}[0]{\underline{G}^{L}}
\newcommand*{\tGL}[0]{\protect\utilde{G}^{L}}

\newcommand*{\tGE}[0]{\protect\utilde{G}^{E}}
\newcommand*{\tGdc}[0]{\protect\utilde{G}^{dc}}
\newcommand*{\IG}[0]{I^{(\Gamma)}}
\newcommand*{\IGiso}[0]{I_{\text{iso}}^{(\Gamma)}}
\newcommand*{\D}[0]{\underline{D}}
\newcommand*{\tD}[0]{\protect\utilde{D}}
\newcommand*{\R}[0]{\vec{R}}
\newcommand*{\Rp}[0]{\vec{R}^{\prime}}
\newcommand*{\kv}[0]{\vec{k}}

\newcommand*{\khat}[0]{\hat{k}}
\newcommand*{\kx}[0]{k_{x}}
\newcommand*{\ky}[0]{k_{y}}
\newcommand*{\kz}[0]{k_{z}}
\newcommand*{\f}[0]{\vec{f}}
\newcommand*{\uv}[0]{\vec{u}}

\begin{document}

\title{Convergence rate for numerical computation of the lattice Green's function}
\author{M. Ghazisaeidi}
\affiliation{Department of Mechanical Science and Engineering, University
of Illinois at Urbana-Champaign, Urbana, Illinois 61801, USA}
\author{D. R. Trinkle}
\affiliation{Department of Materials Science and Engineering, University of
Illinois at Urbana-Champaign, Urbana, Illinois 61801, USA}

\begin{abstract}
Flexible boundary condition methods couple an isolated defect to bulk
through the bulk lattice Green's function.  The inversion of the
force-constant matrix for the lattice Green's function requires Fourier
techniques to project out the singular subspace, corresponding to uniform
displacements and forces for the infinite lattice.  Three different
techniques---relative displacement, elastic Green's function, and
discontinuity correction---have different computational complexity for a
specified numerical error.  We calculate the convergence rates for
elastically isotropic and anisotropic cases and compare them to analytic
results.  Our results confirm that the discontinuity correction is the most
computationally efficient method to compute the lattice Green's function.
\end{abstract}

\maketitle

\section{Introduction}
Atomic-scale simulation of isolated defects with a computationally
tractable number of atoms requires careful choice of boundary
conditions. Periodic or fixed boundary conditions introduce fictitious
forces when relaxing the geometry of defects; reducing the error requires
increasing the number of atoms.  Flexible boundary condition methods avoid
these errors by instead using harmonic lattice response for atoms away from
the defect.  In particular, the bulk lattice Green's function (LGF) gives
the short- and long-range displacements in response to a point or line
force.  Sinclair et~al.\cite{Sinclair:flex} introduced flexible boundary
conditions for studying defects such as cracks\cite{Thomson:rel,crack},
dislocations\cite{Rao-dislocation,Rao-screw,Yang-screw,Rao-slip}, vacancies
and free surfaces\cite{Tewary:surf} with classical potentials and isolated
screw or edge dislocations with density-functional
theory\cite{Rao-screw-DFT,ww:flex,ww-screw-DFT,al-prl}.  Evaluation of the
LGF in real space involves the inverse Fourier transform of a function with
a singularity at the $\Gamma$-point ($k=0$), which requires algorithmic
approaches to evaluate numerically.  Relative displacement
method\cite{Thomson:rel,OHS}, elastic Green's function (EGF)
correction\cite{Tewary:SC} and discontinuity correction\cite{lgf} are three
techniques to numerically evaluating the bulk LGF.  We compare all three
methods to determine the most computationally efficient
approach. Section~\ref{sec:background} defines the harmonic response
functions, the relative displacement method, elastic Green's function
correction and discontinuity correction, and the convergence evaluation
methodology.  Section~\ref{sec:results} follows with the convergence
results and discussion.  We find that the discontinuity correction has the
fastest convergence rate over the relative displacement and elastic Green's
function correction and verify our predicted convergence rates with a
simple model and density functional theory results for Al.

\section{Background}
\label{sec:background}
The lattice Green's function $\GL(\R-\Rp)$ relates the displacements
$\uv(\R)$ of atom $\R$ to the internal forces $\f(\Rp)$ on another atom
$\Rp$ of the lattice through
\begin{equation}
       \uv(\R)=\sum_{\Rp}\GL(\R-\Rp)\f(\Rp).
\label{eqn:GFdef}
\end{equation}
Conversely, the forces on an atom can be expressed in terms of
displacements through the force constant matrix $\D(\R-\Rp)$ by
\begin{equation}
   \f(\R)=-\sum_{\Rp} \D(\R-\Rp) \uv(\Rp).
\label{eqn:Ddef}
\end{equation}
Translational invariance of an infinite lattice makes $\GL$ a function of
the relative positions of two atoms.  Substituting eqn.~(\ref{eqn:Ddef})
into eqn.~(\ref{eqn:GFdef}) gives
$\sum_{\Rp}\GL(\R-\Rp)\D(\Rp)=-\textbf{1}\delta(\R)$, where
$\delta\left(\vec{R}\right)$ is the Kronecker delta function.  A constant
shift in the atoms positions does not produce internal forces, giving the
sum rule $ \sum_{\R}{\D(\R)}=0$ and making $\GL(\R)$ the pseudoinverse of
$\D(\R)$ in the subspace without uniform displacements or forces.  Fourier
transform of the lattice functions are defined as
\begin{equation*}
  \tGL(\kv)=\sum_{\R}e^{i\kv\cdot\R}\GL(\R),\quad
  \GL(\R)=\int_{BZ}\frac{d^{3}k}{(2\pi)^{3}}e^{-i\kv\cdot\R}\tGL(\kv).
\end{equation*}
for $\kv$ in the Brillouin zone (BZ). The integral can be approximated by a
discrete sum of $N_{k}$ points as
$\GL(\R)=\frac{1}{N_{k}}\sum_{\kv}e^{-i\kv\cdot\R}\tGL(\kv)$.  In
reciprocal space, the matrix inverse relation and the sum rule are $
\tGL(\kv)\tD(\kv)=1$ and $\tD(\vec{0})=0$ respectively.  For a single atom
crystal basis, $\tD(\kv)$ expands as for small $\kv$ as
\begin{equation}
\tD(\kv)=\sum_{\R}\D(\R)[1-\frac{(\kv\cdot\R)^{2}}{2!}+\cdot\cdot\cdot]\simeq
-\frac{1}{2}\sum_{\R}{(\kv\cdot\R)^{2}}\D(\R),
\label{eqn:DK}
\end{equation}
due to the inversion symmetry of $\D(\R)$.  At the $\Gamma$-point,
$\tD(\kv)$ is of the order $k^{2}$, so $\tGL(\kv)$ has a second order pole.
The discrete inverse Fourier transform of $\tGL(\kv)$ does not converge due
to this singularity.

\begin{figure*}
 \begin{center}
 \includegraphics[width=6.5 in]{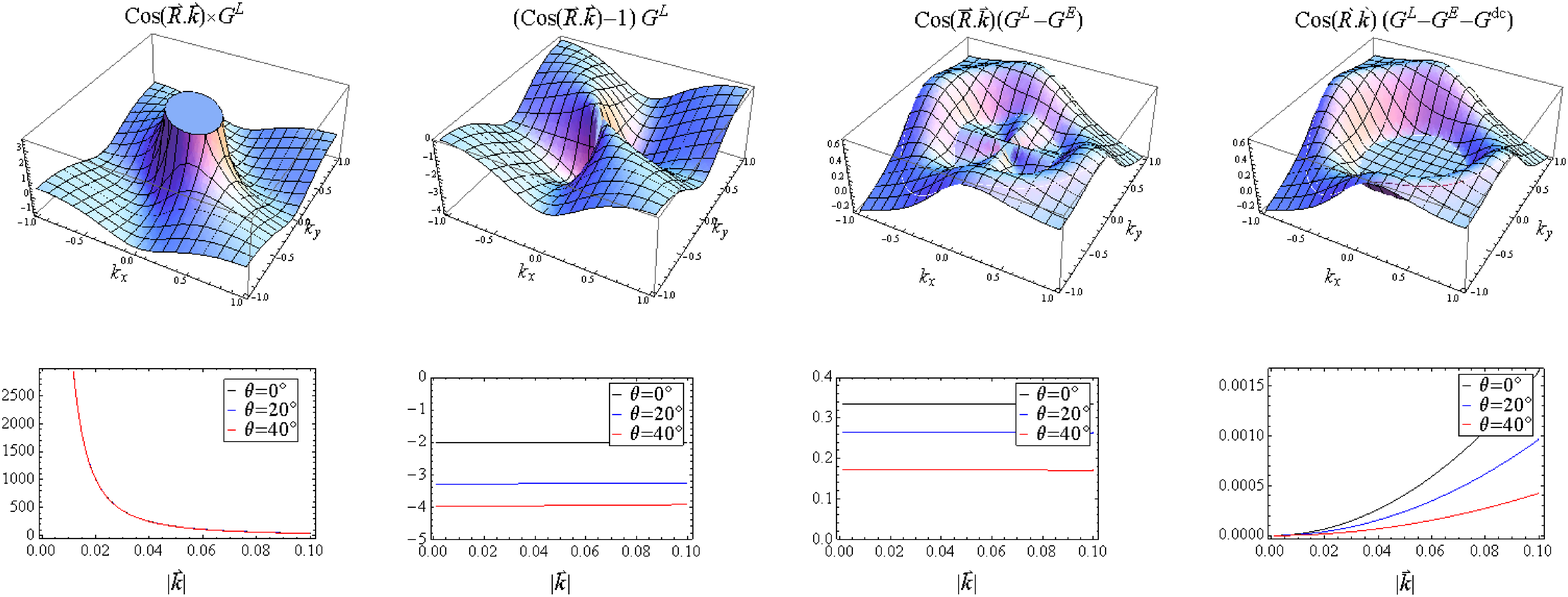}
  
   Lattice Green's function \hspace{0.4cm}  Relative displacement\hspace{0.5cm} EGF correction \hspace{0.5cm} Discontinuity correction \\
 \end{center}

\caption{Integrand of the inverse Fourier transform for LGF, relative
  displacement method, EGF correction and discontinuity correction at
  $\R=(1,1)$.  $\tGL(\kv)$ has a second order pole at the $\Gamma$-point.
  The relative displacement method avoids the pole by considering only the
  displacements relative to a fixed point.  The EGF correction removes the
  second order pole by subtracting a cutoff elastic Green's function.
  Removal of the pole creates a discontinuity independent of
  $\left|\vec{k}\right|$ at the $\Gamma$-point.  The discontinuity
  correction removes the discontinuity created by EGF correction.  The
  remaining part of the integrand is smooth in the entire Brillouin Zone.
  The bottom row shows the variation of the integrand as a function of
  $\left|\vec{k}\right|$ when the origin is approached from different
  angles $\theta=\tan^{-1}(k_{y}/k_{x})$.  A discontinuity at the
  $\Gamma$-point is created by the relative displacement method and EGF
  correction where the integrand is independent of $\left|\vec{k}\right|$
  but depends on the direction of approaching the origin.  Notice the
  difference in vertical scale for the LGF---which has a pole---from
  relative displacement and EGF correction---which have a
  discontinuity---and the discontinuity correction---which approaches zero
  quadratically.}
\label{fig:integrand}
\end{figure*}

Fig.~\ref{fig:integrand} shows the relative displacement method, elastic
Green's function correction and discontinuity correction which are used to
avoid the singularity in LGF.  {\bf Relative
displacement.}\cite{Thomson:rel,OHS} Rigid body translations leave the
potential energy of the lattice unchanged, so it is enough to calculate
only the relative displacements of atoms. Choosing an arbitrary atom as an
undisplaced origin requires calculation of $\GL(\R)-\GL(\vec{0})
=\int\frac{d\kv}{(2\pi)^{3}}\tGL(\kv)e^{i\kv\cdot\R}
-\int\frac{d\kv}{(2\pi)^{3}}\tGL(\kv)e^{i\kv\cdot\vec{0}}$ which reduces to
\begin{equation}
  \GL(\R)-\GL(\vec{0})= \int\tGL(\kv)(\cos(\kv\cdot\R)-1)d\kv,
\label{eqn:rel}
\end{equation}
due to the sum rule and inversion symmetry.  For small $k$,
$\cos(\kv\cdot\R)-1$ is of the order $k^{2}$ which cancels out the second
order pole in $\tGL(\kv)$ leaving a $\kv$-direction dependent discontinuity
at the $\Gamma$-point.  The discretized version of eqn.~(\ref{eqn:rel}) is
$\frac{1}{N_{k}}\sum_{\kv}(\cos(\kv\cdot\R)-1)\tGL(\kv)$.  {\bf Elastic
Green's function correction.} Following the procedure and notations of [12]
and using eqn.~(\ref{eqn:DK}), $\tGL(\kv)$ for small $k$ expands as
\begin{eqnarray*}
\tGL(\kv) &=&[\tD(\kv)]^{-1} \\
    &=  & [k^{2}\tilde{\Lambda}^{(2)}(\hat{k})-k^{4}\tilde{\Lambda}^{(4)}(\hat{k})+O(k^{6})]^{-1} \\
    & = & k^{-2}[\tilde{\Lambda}^{(2)}(\hat{k})]^{-1}+[\tilde{\Lambda}^{(2)}(\hat{k})]^{-1}\tilde{\Lambda}^{(4)}(\hat{k})[\tilde{\Lambda}^{(2)}(\hat{k})]^{-1}+O(k^{-2}) \\
     &=& \tGE(\kv)+\tGdc(\kv)+O(k^{2}),
\end{eqnarray*}
where $k^2\tilde{\Lambda}^{(2)}(\hat{k})$ and
$k^4\tilde{\Lambda}^{(4)}(\hat{k})$ are the second and fourth order terms
in a small $k$ expansion of $\tD(\kv)$. The Fourier transform of the
elastic Green's function $\tGE(\kv)$ is the second order pole and
$\tGdc(\kv)$ is a $\kv$-direction dependent discontinuity~\cite{lgf}.  The
lattice Green's function can be separated into an elastic part $\tGE(\kv)$
which should be inverse Fourier transformed analytically and the rest which
no longer has a pole (it still has a discontinuity). The remaining part of
the LGF can be inverse transformed numerically by
$\frac{1}{N_{k}}\sum_{k}\cos(\kv\cdot\R)(\tGL(\kv)-\tGE(\kv) f_{cut}(\kv))$
where $f_{cut}$ is a cutoff function that smoothly vanishes on the
Brillouin zone edges.  Removal of the second order pole by subtraction of a
cutoff version of elastic Green's function is used in the semicontinuum
method of Tewary\cite{Tewary:SC}.  {\bf Discontinuity correction.} To
further improve convergence, the discontinuity correction treats the
$\tGdc(\kv)$ part analytically\cite{lgf}.  In this case, the remaining
portion of $\tGL(\kv)$ given by
$\frac{1}{N_{k}}\sum_{k}\cos(\kv\cdot\R)(\tGL(\kv)-(\tGE(\kv)+\tGdc(\kv))
f_{cut}(\kv))$ is smooth and can be integrated numerically more
efficiently.

We expect the convergence rate of the discontinuity correction method to be
consistent with the results for integration of smooth periodic functions,
while the convergence of relative displacement and elastic Green function
correction methods should be dominated by the discontinuity.  With
$N_{\text{div}}$ denoting the number of partitions in each direction,
mid-point rule gives a $N_{\text{div}}^{-4}$ scale for convergence rate of
such integrals in all dimensions\cite{INT,NUM}.  The number of $k$ points
$N_{k}$ is $N_{\text{div}}^{d}$ for dimensionality $d=1,2,3$; therefore,
the convergence rate of the mid point rule scales as $N_{k}^{-4/d}$.
However the EGF correction and relative displacement method should have a
poorer convergence compared to discontinuity correction due to the
discontinuity that they create at the $\Gamma$-point.  Since the integrand
is smooth elsewhere, we expect the error to be dominated by the area/volume
around $\Gamma$-point and therefore be of the order of $N_{k}^{-1}$ or
$N_{\text{div}}^{-d}$.

We check the predictions of convergence for the three methods using (1) a
simple cubic nearest neighbor model and (2) fcc Al.  First, as a simplified
case we consider a square (cubic in 3D) elastically isotropic lattice with
nearest neighbor interactions and lattice constant $a_{0}=\pi$. We consider
only one component of the full matrix: $G^{L}(k_{x},k_{y})=[\sin^2(\pi
k_{x}/2)+\sin^2(\pi k_{y}/2)]^{-1}$.  The second order pole is given by the
elastic Green's function $G^{E}(\kx,\ky)=\frac{4}{\pi
\left|\kv\right|^{2}}$ which is multiplied by a cutoff function to vanish
smoothly at the BZ edges.  The discontinuity correction is given by
$G^{dc}(k_{x},k_{y})=\frac{k_{x}^{4}+k_{y}^{4}}{3\left|\vec{k}\right|^{4}}$
which is also multiplied by the cutoff function.  In three dimensions we
have
\begin{eqnarray*}
 G^{L}(\kx,\ky,\kz)&=&[\sin^2(\pi\kx/2)+\sin^2(\pi\ky/2)+\sin^2(\pi\kz/2)]^{-1}\\ 
 G^{E}(\kx,\ky,\kz) & = & \frac{4}{\pi \left|\vec{k}\right|^{2}}\\
 G^{dc}(\kx,\ky,\kz) & = & \frac{\kx^{4}+\ky^{4}+\kz^{4}}{3\left|\vec{k}\right|^{4}}.
\end{eqnarray*}
For our elastically anisotropic Al lattice, we obtain the force constant
matrix $\D(\R)$ from DFT using ultrasoft pseudopotentials with the
generalized gradient approximation\cite{al-prl}.  The numerical integration
over the BZ is done with a uniform mesh evaluating the integrand at the mid
points. Even and odd values of $N_{\text{div}}$ give meshes that include or
avoid the $\Gamma$-point---what we call $\Gamma$ and non-$\Gamma$ centered
meshes respectively.  When applying the relative displacement method and
EGF correction, the value of the integrand--- which is discontinuous at
$k=0$--- is assigned zero at the $\Gamma$-point. We calculate the numerical
error as a function of $N_{k}$ and $N_{\text{div}}$ to compare the
efficiency of the three methods.

\section{Results and Discussion}
\label{sec:results}

\begin{figure}
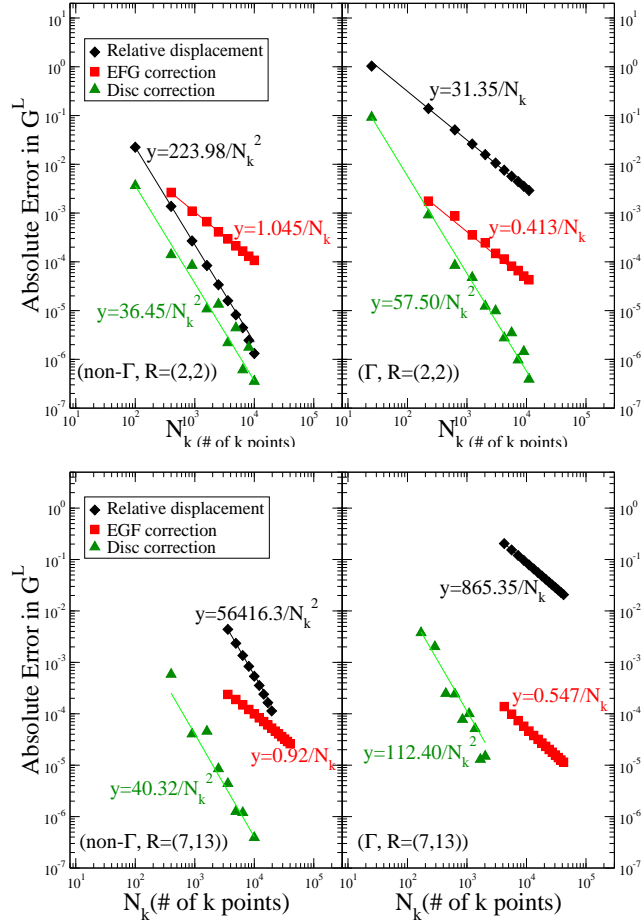

  \begin{center}
    \includegraphics[width=3.3 in,height=60mm]{fig2a.eps}\\
    \vspace{2mm}
    \includegraphics[width=3.3 in,height=60mm]{fig2b.eps}
  \end{center}

\caption{Convergence rate with number of k-points of the relative
  displacement method, EGF correction and discontinuity correction in a 2D
  square lattice. We expect $N_{k}^{-2}$ convergence for discontinuity
  correction and poorer $N_{k}^{-1}$ convergence for EGF correction and
  relative displacement method.  Using a non-$\Gamma$ centered mesh (left)
  causes an unusually fast convergence for the relative displacement method
  in elastically isotropic materials.  The exponents in the power law
  scalings are not affected by the value of $\R$, while prefactors are
  changed in relative displacement method and are of the same order in EGF
  and discontinuity corrections.}
\label{fig:2dlatt}
\end{figure}

Fig.~\ref{fig:2dlatt} shows the convergence rates of relative displacement
method, EGF correction and discontinuity correction in the square lattice
case.  The discontinuity correction and EGF correction scale as
$N_{k}^{-2}$ and $N_{k}^{-1}$ respectively as expected.  The value of
$\vec{R}$ does not affect the power law scalings of the convergence. The
prefactors on the other hand, are changed in the relative displacement
method and are of the same order in EGF and discontinuity corrections.
While the $N_{k}^{-1}$ convergence for relative displacement method
obtained by a $\Gamma$ centered mesh is in accordance with the analytical
predictions, use of a non-$\Gamma$ centered mesh produces a convergence
faster than expected for this method. This is an artifact of the isotropy
of the EGF.

The integrand in the relative displacement method is
$I(\kv)=(\cos(\vec{k}\cdot\vec{R})-1)\tGL(\kv)$.  Near the $\Gamma$-point,
$\tGL(\kv)$ matches $\tGE(\kv)$ and the leading term in the integrand is
\begin{equation*}
\IG(\kv)= - k^{2}R^{2}(\hat{k}\cdot\hat{R})^{2}\frac{\tGE(\khat)}{2k^{2}}.
\end{equation*}
For an isotropic EGF, $\tGE(\khat)$ is constant, so 
\begin{equation*}
\IGiso(\kv)=-\frac{1}{2}\tGE\cdot R^{2}\cos^2(\theta_{\khat,\hat{R}})
\end{equation*}
where $\theta_{\khat,\hat{R}}$ is the angle between vectors $\khat$ and
$\hat{R}$.  The value of the integral over a square $k_{0}\times k_{0}$
region around $k=0$, for small $k$ is
\begin{equation}
\int_{k_0^2}\IGiso(\kv)d^{2}k=
\int_{-k_{0}/2}^{k_{0}/2}\int_{-k_{0}/2}^{k_{0}/2}
(-\frac{\tGE}{2}R^{2}\cos^2(\theta)) d\kx d\ky=
-\frac{k_0^2}{4}\tGE R^{2}.
\label{eqn:exact}
\end{equation}
The midpoint rule integration of the same region with a non-$\Gamma$
centered mesh uses the k points $\kv_{1}=(k_0/2,k_0/2)$,
$\kv_{2}=(-k_0/2,k_0/2)$, $\kv_{3}=(-k_0/2,-k_0/2)$ and
$\kv_{4}=(k_0/2,-k_0/2)$ each contributing area $k_0^2/4$.  The angle
between each $\kv_{i}$ and $\hat{R}$, are $\theta_{1}$,
$\theta_{2}=\theta_{1}+\pi/2$, $\theta_{3}=\theta_{1}+\pi$ and
$\theta_{4}=\theta_{1}+3\pi/2$. Therefore, the numerical approximation for
the integral around the $\Gamma$-point is
\begin{equation}
\begin{split}
\bar{I}_\text{iso}&=\frac{k_0^2}{4}
\left[\IGiso(\kv_{1})+\IGiso(\kv_{2})+\IGiso(\kv_{3})+\IGiso(\kv_{4})\right]\\
&=-\frac{k_0^2}{8}\tGE
R^2\left[\cos^{2}\theta_{1}+\sin^{2}\theta_{1}+\cos^{2}\theta_{1}+\sin^{2}\theta_{1}\right]
=-\frac{k_0^2}{4}\tGE R^{2}.
\end{split}
\label{eqn:nongamma}
\end{equation}
which is equal to the exact value of the integral around $\Gamma$-point
given by eqn.~(\ref{eqn:exact}).  To avoid the effect of the discontinuity
at the origin using a $\Gamma$ centered mesh, the $\Gamma$-point
contribution to the integral is considered zero while its actual value is
given by eqn.~(\ref{eqn:exact}).  This is the source of the dominant error
in relative displacement method on a $\Gamma$ centered mesh which according
to eqn.~(\ref{eqn:exact}) accounts for the $R$ dependence of the error. The
$R$ dependence of the error is verified by comparing the ratio of
prefactors of the relative displacement convergence laws for different $R$
values and the corresponding $R^{2}$ in Fig.~\ref{fig:2dlatt} which are
both approximately 27.  On the other hand, the non-$\Gamma$ centered mesh
automatically gives the exact value of the integral around the origin based
on eqn.~(\ref{eqn:nongamma}) and thus produces a faster convergence limited
only by the convergence of smooth periodic functions.  Note that if
$\tGE(\khat)$ depends on $\khat$---which is the case for anisotropic
elastic response---the numerical approximation of the integral around
$\Gamma$ will not be equal to its exact value. Therefore, in general
anisotropic problems the non-$\Gamma$ mesh is not special.

\begin{figure}
 \begin{center}
   \includegraphics[width=3.3 in]{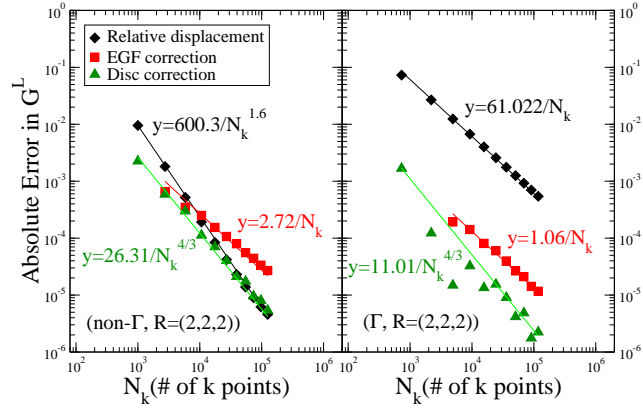}
 \end{center}

\caption{Convergence rate with number of k-points of the relative
  displacement method, elastic GF correction and discontinuity correction
  in a 3D cubic lattice.  The error for discontinuity correction method
  scales as $N_{k}^{-4/d}$ where the dimension $d$ is equal
  to three here.  Note that using a non-$\Gamma$ centered mesh creates a faster
  convergence for the relative displacement method as observed in the 2D
  case.}
\label{fig:3dlatt}
\end{figure}

Fig.~\ref{fig:3dlatt} shows that the 3D results follow the same trend as
the 2D ones in accordance with the expected values. Both $\Gamma$ centered
and non-$\Gamma$ centered meshes give $N_{k}^{-4/d}$ ($d=3$) and
$N_{k}^{-1}$ scale for the convergence rate of discontinuity correction and
EGF correction respectively. Similar to the trend observed in 2D case the
$\Gamma$ centered mesh produces the expected $N_{k}^{-1}$ scale for the
convergence of relative displacement method and the non-$\Gamma$ centered
mesh produces faster convergence due to the isotropy of the elastic Green's
function.

\begin{figure}
 \begin{center}
   \includegraphics[width=3.3 in]{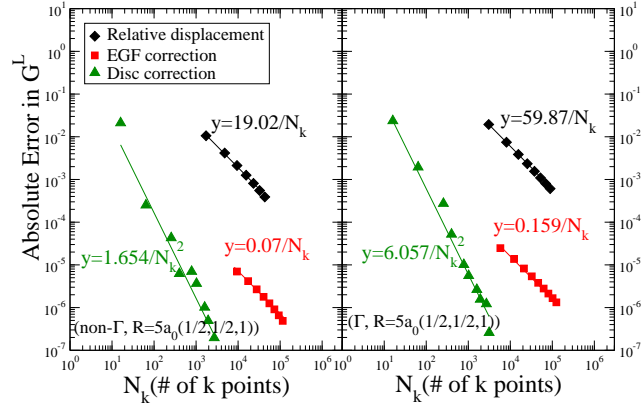}
 \end{center}

\caption{ Convergence rate with number of k-points of the relative
  displacement method, elastic Green's function correction and
  discontinuity correction in computation of the $G_{11}$ component of a 2D
  LGF in Al. The Al lattice constant $a_{0}$ is 4.04\AA.
  The convergence trend of LGF calculations in a FCC lattice is the same as
  the one observed in the simplified problem which is also consistent with
  the expected values. Note that use of the non-$\Gamma$ centered mesh does
  not cause a fast convergence for relative displacement method due to the
  anisotropy of the elastic Green's function.}
\label{fig:Al}
\end{figure}

Fig.~\ref{fig:Al} shows that the convergence trends are not changed for an
anisotropic long range interaction---fcc Al---except for relative
displacement method. The convergence rates of the three methods for the two
dimensional LGF are shown.  This is a relevant case that occurs in modeling
dislocations. The lattice is periodic in the threading direction $[110]$
which is appropriate for studying screw dislocations.  With a non-$\Gamma$
centered mesh, the anisotropy of the elastic Green's function eliminates
the fast convergence of the relative displacement method.  The convergence
trends of the three methods show that these trends are not specific to the
simplifying assumptions of isotropy or short range interactions and
therefore can be trusted in realistic calculations.

\begin{table}
\caption{Effect of dimension on the convergence rate with number of
  k-points and number of divisions for the relative displacement method,
  EGF correction and discontinuity correction.  $N_{\text{div}}$ is proportional
  to $1/h$ , the inverse grid spacing and $N_{k}=N_{\text{div}}^{d}$.  The
  discontinuity correction scales as $N_{\text{div}}^{-4}$ (or $N_{k}^{-4/d}$)
  while the EGF correction and relative displacement method scale as
  $N_{k}^{-1}$ (or $N_{\text{div}}^{-d}$).}
\label{tab:dimen}

\begin{center}
\begin{tabular}{lcccc}
       &\hspace{1cm}  2D& &\hspace{1cm}3D& \\ 
\underline{Power law scaling of error with} & \underline{ $N_{k}$} & \underline{$N_{\text{div}}$} & \underline{ $N_{k}$} &\underline{$N_{\text{div}}$} \\
 Disc correction &   $ -2$ & $ -4$ & $-4/3$ & $-4$ \\
 
 EGF correction &    $-1$ & $-2$ & $-1$ & $-3$ \\
 
 Rel.~displacement &    $-1$ & $-2$ &  $-1$ & $-3$\\
%\hline
\end{tabular}
\end{center}
\end{table}

Table~\ref{tab:dimen} summarizes the convergence results for the three
methods.  The expected convergence rate for a numerical integral of a
smooth periodic function evaluated by mid-point rule is
$N_{\text{div}}^{-4}$.  When expressed in terms of the number of $k$-points
used in evaluating the integral $N_{k}$, the convergence rate would be
proportional to $N_{k}^{-4/d}$. Since the discontinuity correction leaves a
smooth periodic part of the integrand, it follows the above convergence
rate.  The EGF correction and relative displacement method also converge
with the scale of $N_{k}^{-1}$ or $N_{\text{div}}^{-d}$. Therefore the
discontinuity correction method has the fastest convergence rate. The
convergence rates imply that a certain amount of error is achieved with
less $N_{k}$ by discontinuity correction method compared to EGF correction
or relative displacement method which means that the discontinuity
correction requires the least computational effort. Although the EGF
correction and relative displacement method require comparable
computational effort, the $R$ dependence of the prefactors suggests that
the relative displacement method takes even more $k$-points than the EGF
correction. Also note that there is a trade-off between less computational
effort and more complex algorithms.  EGF and discontinuity corrections
calculate the elastic Green's function and discontinuity correction parts
of the LGF analytically while relative displacement method does not require
additional analytic evaluations.

\section{Conclusion}
We find the most efficient method to compute the lattice Green's function
to be the discontinuity correction. The relative displacement method,
elastic Green's function correction and discontinuity correction have all
been used in different calculations; we applied and compared the three
methods to calculate LGF for a square nearest neighbor lattice and long
range fcc Al. The convergence trends match the analytical values with an
unusual exception for lattices with isotropic elastic Green's function.  It
is shown that the discontinuity correction improves the convergence rate to
quadratic convergence for 2D calculations compared to linear convergence
for the relative displacement and elastic Green's function correction.

\end{document}